\documentclass{article}
\usepackage{spconf,amsmath}

\usepackage{multirow}
\usepackage{graphicx}
\usepackage{amsmath}
\usepackage[psamsfonts]{amssymb}
\usepackage{amsxtra}
\usepackage{threeparttable}
\usepackage{amsfonts}
\usepackage{mathrsfs}
\usepackage{arydshln}
\usepackage{cite}
\usepackage{enumitem}
\usepackage{url}

\makeatletter
\let\MYcaption\@makecaption
\makeatother
\usepackage{subcaption}
\makeatletter
\let\@makecaption\MYcaption
\makeatother
\makeatletter
\renewcommand\section{\@startsection{section}{1}{\z@}
                      {0.5ex \@plus 0ex \@minus -2ex}
                      {0.5ex \@plus 0ex}
                      {\normalfont\Large\bfseries}}
\renewcommand\subsection{\@startsection{subsection}{2}{\z@}
                      {0.5ex \@plus 0ex \@minus -2ex}
                      {0.5ex \@plus 0ex}
                      {\normalfont\large\bfseries}}
\renewcommand\subsubsection{\@startsection{subsubsection}{3}{\z@}
                      {0.5ex \@plus 0ex \@minus -2ex}
                      {0.5ex \@plus 0ex}
                      {\normalfont\normalsize\bfseries}}
\def\@listi{\leftmargin\leftmargini
            \parsep 1.0pt
            \topsep 0.2\baselineskip \@minus 0.1\baselineskip
            \itemsep 1.0pt \relax}
\let\@listI\@listi
\makeatother

\newcounter{num}
\newcommand{\rnum}[1]{\setcounter{num}{#1} \roman{num}}

\title{Convolutional Neural Networks
Considering Local and Global features for Image Enhancement}
\name{Yuma Kinoshita and Hitoshi Kiya}
\address{Tokyo Metropolitan University, Tokyo, Japan}
\begin{document}\sloppy
\setlength{\tabcolsep}{1.0pt}
\setlength{\textfloatsep}{2.0pt}
\setlength{\floatsep}{2.0pt}
\setlength{\abovecaptionskip}{10.0pt}
\ninept
\maketitle
\begin{abstract}
  In this paper, we propose a novel convolutional neural network (CNN) architecture considering
  both local and global features for image enhancement.
  Most conventional image enhancement methods,
  including Retinex-based methods,
  cannot restore lost pixel values caused by clipping and quantizing.
  CNN-based methods have recently been proposed to solve the problem,
  but they still have a limited performance
  due to network architectures not handling global features.
  To handle both local and global features,
  the proposed architecture consists of three networks:
  a local encoder, a global encoder, and a decoder.
  In addition, high dynamic range (HDR) images
  are used for generating training data for our networks.
  The use of HDR images makes it possible to train CNNs with better-quality images
  than images directly captured with cameras.
  Experimental results show that the proposed method can produce
  higher-quality images than conventional image enhancement methods
  including CNN-based methods,
  in terms of various objective quality metrics: TMQI, entropy, NIQE, and BRISQUE.
\end{abstract}
\renewcommand{\thefootnote}{\fnsymbol{footnote}}
\footnote[0]{This work was supported by JSPS KAKENHI Grant Number JP18J20326.}
\renewcommand{\thefootnote}{\arabic{footnote}}
\section{Introduction}
  The low dynamic range (LDR) of modern digital cameras
  is a major factor that prevents cameras from capturing images as well as human vision.
  This is due to the limited dynamic range that imaging sensors have,
  resulting in low-contrast images.
  Enhancing such images reveals hidden details.

  Various kinds of research on single-image enhancement have been reported
  \cite{zuiderveld1994contrast, wu2017contrast,
  guo2017lime, fu2016weighted, kinoshita2018automatic_trans}.
  Most image enhancement methods can be divided into two types:
  histogram equalization (HE)-based methods and Retinex-based methods.
  However, both HE- and Retinex-based methods cannot restore lost pixel values due to
  quantizing and clipping.
  The problem leads to banding artifacts in enhanced images.
  For this reason, image enhancement methods based on
  convolutional neural networks (CNNs)
  are expected to solve the problem that traditional methods have.

  CNNs have been successfully used in many image-to-image translation tasks
  such as single image super-resolution, image inpainting, and image segmentation.
  Recently, CNN-based image enhancement methods
  have also been developed
  \cite{chen2018learning, yang2018image, cai2018learning, kinoshita2019image,
  gharbi2017deep, shen2017msrnet}.
  Most of these methods employ a U-Net \cite{ronneberger2015unet}-based
  network architecture.
  However, U-Net cannot extract global features of images when large input images are given
  or when U-Net itself does not have a sufficient number of layers.
  In this case, output images are often distorted because both local and global features
  are needed for image enhancement,
  whereas only local ones are needed for other tasks
  such as super-resolution.

  In this paper, we propose a novel CNN architecture that considers
  local and global features for image enhancement.
  The proposed architecture consists of three networks:
  a local encoder, a global encoder, and a decoder.
  The use of the local encoder and the global encoder enables us to extract
  both local and global features, even when large images are given as inputs.
  In the decoder, these features are combined to produce enhanced images.
  The proposed CNNs can be trained with small images,
  and images of various sizes can be enhanced by the CNNs without distortions.
  In addition, we utilize tone-mapped images from existing high dynamic range (HDR) images
  for the training \cite{reinhard2002photographic, murofushi2013integer}.
  Target LDR images mapped from HDR ones have better quality than
  those directly captured from cameras
  because HDR images contain more information.
  
  We evaluate the effectiveness of the proposed image enhancement network
  in terms of the quality of enhanced images in a number of simulations,
  where the tone mapped image quality index (TMQI), discrete entropy,
  naturalness image quality evaluator (NIQE),
  and blind/referenceless image spatial quality evaluator (BRISQUE)
  are utilized as quality metrics.
  Experimental results show that the proposed method outperforms
  state-of-the-art contrast enhancement methods
  in terms of those quality metrics.
  Furthermore, the proposed method does not cause distortions in output images,
  unlike CNNs that do not consider global features.
  
\section{Related work}
  Many image enhancement methods have been studied
  \cite{zuiderveld1994contrast, wu2017contrast,
  guo2017lime, fu2016weighted,
  kinoshita2018automatic_trans, kinoshita2019scene}.
  Among the methods, HE has received
  the most attention because of its intuitive implementation quality and
  high efficiency.
  It aims to derive a mapping function such that the entropy of
  a distribution of output luminance values can be maximized.
  However, HE often causes over-enhancement.
  To avoid this,
  numerous improved methods based on HE have also been developed
  \cite{zuiderveld1994contrast, wu2017contrast}.
  Another way for enhancing images is to use the Retinex theory \cite{land1977retinex}.
  Retinex-based methods \cite{guo2017lime, fu2016weighted} decompose images into
  reflectance and illumination, and then enhance images by manipulating illumination.
  However, without the use of CNNs, these methods
  cannot restore lost pixel values due to clipping and quantizing.
  
  Recently, a few CNN-based methods were proposed for single image enhancement
  \cite{chen2018learning, yang2018image, cai2018learning, kinoshita2019image,
  gharbi2017deep, shen2017msrnet}.
  Chen's method \cite{chen2018learning} provides high-quality RGB images
  from single raw images taken under low-light conditions,
  but this method cannot be applied to RGB color images that are not raw images.
  Yang's et al. \cite{yang2018image} proposed a method for enhancing RGB images
  by using two CNNs.
  This method generates intermediate HDR
  images from input RGB images,
  and then produces high-quality LDR images.
  However, generating HDR images from single images is a well-known outstanding problem
  \cite{eilertsen2017hdr, kinoshita2017fast}.
  For this reason, the performance of Yang's method is limited
  by the quality of the intermediate HDR images.
  Furthermore, network architectures of existing methods
  \cite{chen2018learning, yang2018image, cai2018learning, kinoshita2019image}
  only handle local image information,
  so output images can be distorted by CNNs
  due to a lack of global image information
  required for image enhancement,
  which is addressed later in this paper.
  Therefore, we aim to improve the performance of image enhancement
  by using a novel network architecture that considers
  both local and global image information.

\section{Proposed method}
\begin{figure}[!t]
    \centering
    \begin{subfigure}[t]{0.75\hsize}
      \centering
      \includegraphics[width=\columnwidth]{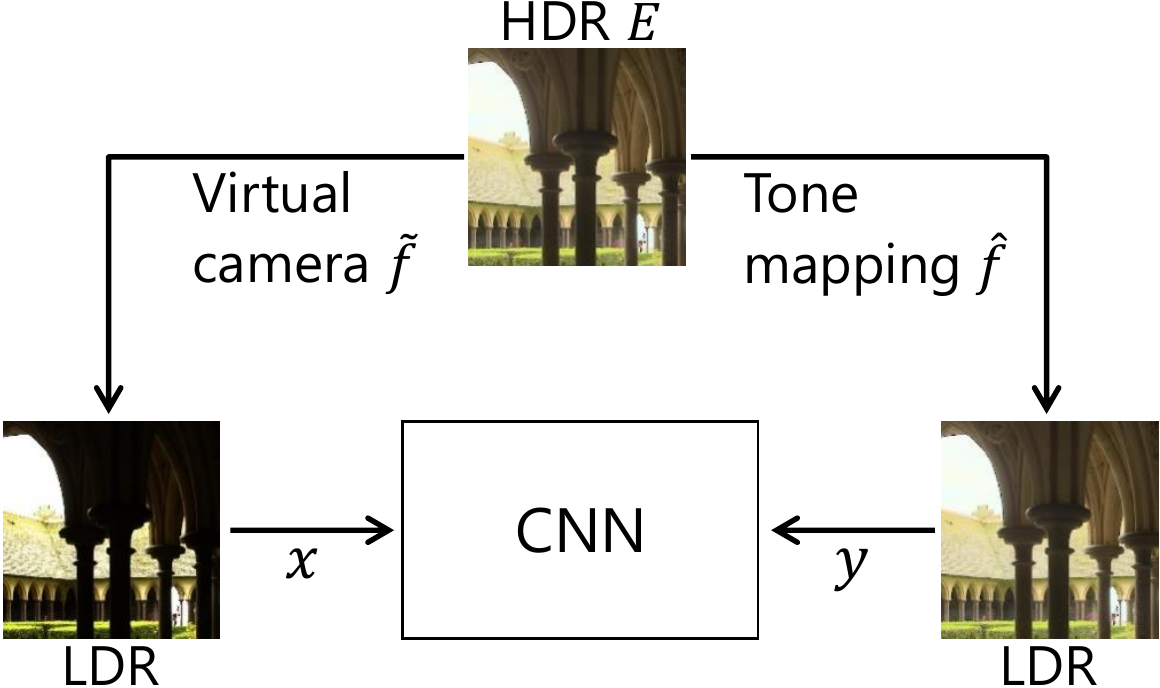}
      \caption{Training \label{fig:train}}
    \end{subfigure}\\
    \centering
    \begin{subfigure}[t]{0.95\hsize}
      \centering
      \includegraphics[width=\columnwidth]{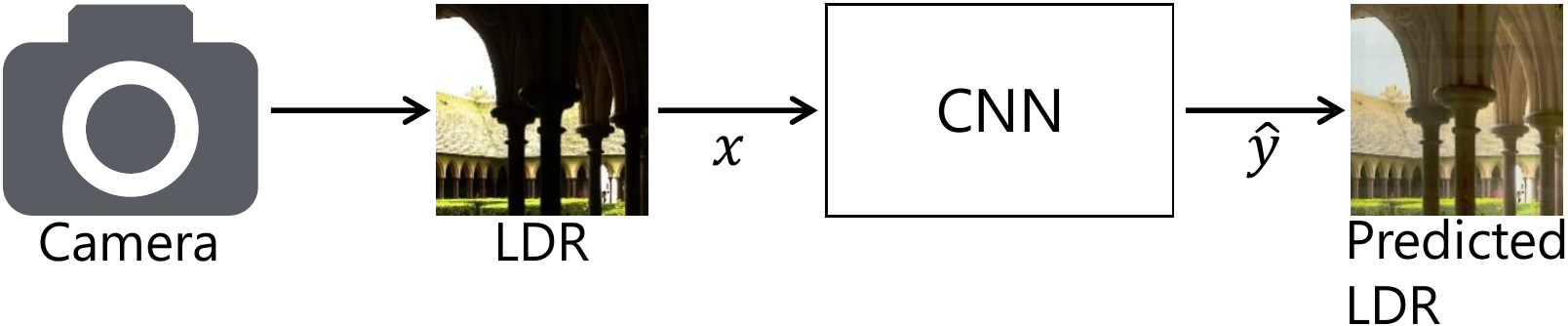}
      \caption{Predicting \label{fig:predict}}
    \end{subfigure}
    \caption{Proposed image enhancement method \label{fig:scheme}}
  \end{figure}

  Figure \ref{fig:scheme} shows an overview of
  our training procedure and predicting procedure.
  In the training, all input LDR images $x$ and target LDR images
  $y$ are generated from HDR images by using
  a virtual camera \cite{eilertsen2017hdr} and tone mapping, respectively.
  This enables us to generate high-quality target images.

  After the training, various LDR images are applied to the proposed CNNs
  as input images,
  where the CNNs then generate high-quality LDR images.
  Detailed training conditions are described in Section \ref{sec:training}.

\subsection{Network architecture}
\begin{figure*}[t]
    \centering
    \includegraphics[clip, width=0.95\hsize]{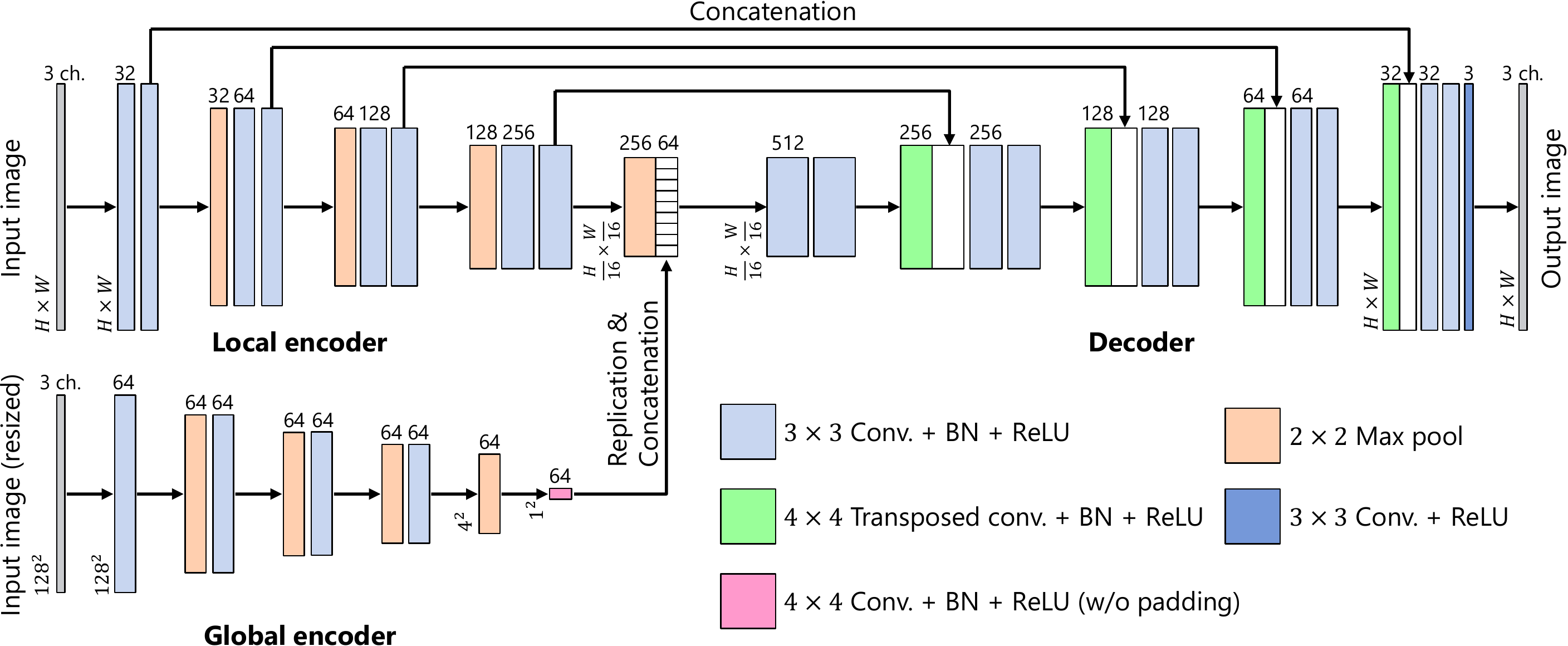}
    \caption{Network architecture.
      Architecture consists of local encoder, global encoder, and decoder.
      Each box denotes multi-channel feature map produced by each layer.
      Number of channels is denoted above each box.
      Feature map resolutions are denoted to the left of a box.
      \label{fig:network}}
  \end{figure*}

  Figure \ref{fig:network} shows the overall network architecture of
  the proposed method.
  The architecture consists of three networks: a local encoder, a global encoder, and a decoder.
  The local encoder and the decoder in the proposed method are almost the same as
  those used in U-Net \cite{ronneberger2015unet}.
  Concatenated skip connections between the local encoder and the decoder is
  also utilized like in U-Net.
  Although U-Net works very well for various image-to-image translation problems,
  its use for image enhancement often causes distortions in output images
  \cite{marnerides2018expandnet}.
  This is due to its network architecture that cannot handle global image information
  (see Section 4).
  For this reason,
  we utilize the global encoder and combine features extracted by both encoders
  to prevent the distortions.
  The input for the local encoder is a $H \times W$ pixels 24-bit color LDR image.
  For the global encoder,
  the input image is resized to a fixed size ($128 \times 128$).

  The proposed CNNs have five types of layers as shown in Fig. \ref{fig:distortion}:
  \begin{description}[style=standard, nosep]
    \item [$3 \times 3$ Conv.+ BN + ReLU]
      which calculates a $3 \times 3$ convolution with a stride of $1$ and a padding of $1$.
      After convolution, batch normalization \cite{ioffe2015batch}
      and the rectified linear unit activation function \cite{glorot2011deep} (ReLU)
      are applied.
      In the local encoder and the decoder, two adjacent 
      $3 \times 3$ Conv.+ BN + ReLU layers will have the same number $K$ of filters.
      From the first two layers to the last ones,
      the number of filters are
      $K = 32$, $64$, $128$, $256$, $512$, $256$, $128$, $64$, and $32$,
      respectively.
      In the global encoder, all layers have $64$ filters.
    \item [$2 \times 2$ Max pool]
      which downsamples feature maps by max pooling with a kernel size of $2 \times 2$ and
      a stride of $2$.
    \item [$4 \times 4$ Transposed Conv. + BN + ReLU]
      which calculates a $4 \times 4$ convolution with a stride of $1/2$ and a padding of $1$.
      After convolution, BN and ReLU are applied.
      From the first layer to the last one
      the numbers of filters are $K = 256$, $128$, $64$, and $32$, respectively.
    \item [$3 \times 3$ Conv. + ReLU]
      which calculates a $3 \times 3$ convolution with a stride of $1$ and a padding of $1$.
      After convolution, ReLU is applied.
      The number of filters in the layer is $3$.
    \item [$4 \times 4$ Conv. + BN + ReLU (w/o padding)]
      which calculates a $4 \times 4$ convolution without padding.
      The number of filters in the layer is $64$.
  \end{description}

\subsection{Training \label{sec:training}}
  Numerous LDR images taken under various conditions, $x$,
  and corresponding high-quality images, $y$, are needed to train
  the CNNs in the proposed method.
  However, collecting a sufficient number is difficult.
  We therefore utilize HDR images $E$ to generate both $x$
  and $y$ by using a virtual camera \cite{eilertsen2017hdr}
  and a tone mapping operation with Mertens's fusion method \cite{mertens2009exposure},
  respectively.
  The use of HDR images also makes it possible to generate better-quality images
  than those directly captured with cameras.
  For training, 831 HDR images were collected from online available databases
  \cite{openexrimage, anyherehdrimage, hdrps, maxplanck, zolliker2013creating,
  nemoto2015visual}.
  
  The training procedure of our CNNs is shown as follows.
  \begin{enumerate}[label=\roman*, align=parleft, leftmargin=*, nosep]
    \item Select 16 HDR images from the 831 HDR images at random.
    \item Generate 16 pairs of an input and target LDR images ($x$, $y$)
      from each HDR image.
      Each pair is generated in accordance with the following steps.
      \begin{enumerate}[label=(\alph*), align=parleft, leftmargin=*, nosep]
        \item Crop HDR image $E$ to an image patch $\tilde{E}$ at $N \times N$ pixels.
          The size $N$ is given as a product of a uniform random number
          in the range $[0.2, 0.6]$
          and the length of the short side of $E$.
          In addition, the position of the patch in $E$
          is also determined at random.
        \item Resize $\tilde{E}$ to $256 \times 256$ pixels.
        \item Flip $\tilde{E}$ horizontally or vertically with a probability of 0.5.
        \item Calculate exposure $X$ from $\tilde{E}$ by
          $X_{i, j} = \Delta t (v) \cdot \tilde{E}_{i, j}$,
          where $(i, j)$ denotes a pixel.
          Shutter speed $\Delta t$ is calculated as
          $\Delta t (v) = 0.18 \cdot 2^v / G(\tilde{E})$ as in \cite{reinhard2002photographic}
          by using a uniform random number $v$ in the range $[-4, 0]$.
          $G(\tilde{E})$ is the geometric mean of luminance of $\tilde{E}$,
        \item Generate an input LDR image $x$ from $X$
          by using a virtual camera $\tilde{f}$, as
          \begin{equation}
            x_{i, j} = \tilde{f}(X_{i, j})
                     = \min
                       \left((1+\eta)\frac{X_{i, j}^\gamma}{X_{i, j}^\gamma + \eta},
                       1 \right),
            \label{eq:virtual_camera}
          \end{equation}
          where $\eta$ and $\gamma$ are random numbers
          that follow normal distributions with a mean of $0.6$ and a variance of $0.1$
          and with a mean of $0.9$ and a variance of $0.1$, respectively.
          Note that eq. (\ref{eq:virtual_camera}) is applied to the luminance of $X$,
          and RGB pixel values of $x$ are then obtained
          so that the RGB value ratios of $x$ are equal to those of $X$.
        \item Generate a target LDR image $y$ from $\tilde{E}$
          by using a tone mapping operator $\hat{f}$ as
          $y = \hat{f}(\tilde{E})$.
          Here, Mertens's multi exposure fusion method \cite{mertens2009exposure}
          is used in the tone mapping operator:
          $\hat{f}(\tilde{E}) = \mathcal{F}(Y^{(-2)}, Y^{(0)}, Y^{(2)})$,
          where,
          similarly to $X$, exposure $Y^{(u)}_{i, j}$ is given as
          $Y^{(u)}_{i, j} = \Delta t (u) \tilde{E}_{i, j}$,
          and $\mathcal{F}(Y^{(-2)}, Y^{(0)}, Y^{(2)})$ indicates
          a function for fusing the three exposures by Mertens's method.
      \end{enumerate}
    \item Predict 16 LDR images $\hat{y}$ from 16 input LDR images $x$ by using the CNNs.
    \item Evaluate errors between predicted images $\hat{y}$ and target images $y$
      by using the mean squared error.
    \item Update filter weights $\omega$ and biases $b$ in the CNNs by back-propagation.
  \end{enumerate}

  In our experiments, the CNNs were trained with 500 epochs,
  where the above procedure was repeated 51 times in each epoch.
  In addition, each HDR image had only one chance to be selected,
  in Step \rnum{1} in each epoch.
  He's method \cite{he2015delving}
  was used for initializing of the CNNs.
  In addition, the Adam optimizer \cite{kingma2014adam} was utilized for optimization,
  where parameters in Adam were set as $\alpha=0.002, \beta_1=0.9$, and $\beta_2=0.999$.

\section{Simulation}
  We evaluated the effectiveness of the proposed method
  by using four objective quality metrics.

\subsection{Simulation conditions}
  In this experiment, test LDR images were generated
  from 283 HDR images that were not used for training,
  in accordance with Steps \rnum{2}(a)--(e).
  In addition, test LDR images were resized to $512 \times 512$ pixels
  in Step \rnum{2}(b).

  The quality of LDR images $\hat{y}$ generated by the proposed method
  was evaluated by four objective quality metrics:
  the tone mapped image quality index (TMQI) \cite{yeganeh2013objective},
  discrete entropy,
  the naturalness image quality evaluator (NIQE) \cite{mittal2013making},
  the blind/referenceless image spatial quality evaluator (BRISQUE)
  \cite{mittal2012no},
  where the original HDR image $\tilde{E}$ was utilized as a reference for TMQI,
  and a database \cite{livehdr, kundu2017large} was used to build a model in BRISQUE.

  The proposed method was compared with five conventional methods:
  histogram equalization (HE),
  contrast-accumulated histogram equalization (CACHE) \cite{wu2017contrast},
  simultaneous reflectance and illumination estimation (SRIE) \cite{fu2016weighted},
  low-light image enhancement via illumination map estimation (LIME) \cite{guo2017lime},
  and deep reciprocating HDR transformation
  \footnote{An approximate implementation
  at \url{https://github.com/ybsong00/DRHT} was utilized}
  (DRHT) \cite{yang2018image},
  where SRIE and LIME are Retinex-based methods and DRHT is a CNN-based one.
\subsection{Results}
  Table \ref{tab:scores} illustrates the average scores of the objective assessment
  for 283 images,
  in terms of TMQI, entropy, NIQE, and BRISQUE.
  In the case of TMQI, a larger value means a higher similarity between a target LDR image
  and an original HDR image.
  A larger value for entropy and BRISQUE indicates that a target LDR image has higher
  quality.
  By contrast, a smaller value for NIQE indicates that a target LDR image has
  less distortions such as noise or blur.
  As shown in Table \ref{tab:scores},
  the proposed method provided the highest average scores
  for entropy, NIQE, and BRISQUE, in the six methods.
  HE also provided high TMQI scores,
  but HE cannot restore lost pixel values and is well known to often causes over-enhancement
  in bright areas \cite{kinoshita2019image}.
  For these reasons, the proposed method outperforms the conventional methods
  in terms of the four metrics.
  \begin{table}[!t]
    \centering
    \caption{Average scores of objective quality metrics}
    {
    \scriptsize
    \begin{tabular}{l|c|r|r|r|r|r|r} \hline\hline
      \multirow{2}{*}{Method} & \multirow{2}{*}{Input}
       & \multicolumn{2}{c|}{HE based} & \multicolumn{2}{c|}{Retinex based}
       & \multicolumn{2}{c}{CNN based} \\ \cline{3-8}
       & & \multicolumn{1}{c|}{HE}
       & \multicolumn{1}{c|}{CACHE \cite{wu2017contrast}}
       & \multicolumn{1}{c|}{SRIE \cite{fu2016weighted}}
       & \multicolumn{1}{c|}{LIME \cite{guo2017lime}}
       & \multicolumn{1}{c|}{DRHT \cite{yang2018image}}
       & \multicolumn{1}{c}{Proposed} \\ \hline
      TMQI & \multicolumn{1}{r|}{0.7547} & \textbf{0.8834} & 0.8531
       & 0.8109 & 0.8407 & 0.7800 & 0.8556 \\
      Entropy & \multicolumn{1}{r|}{3.7861} & 6.0027 & 6.3761
       & 5.3209 & 6.3984 & 5.0554 & \textbf{6.4126} \\
      NIQE & \multicolumn{1}{r|}{5.2876} & 5.5537 & 5.0932
       & 4.8992 & 5.0704 & 9.5137 & \textbf{4.7207} \\
      BRISQUE & \multicolumn{1}{r|}{43.3178} & 45.4301 & 45.4564
       & 45.2134 & 45.3957 & 44.6974 & \textbf{45.7993} \\
      \hline\hline
    \end{tabular}
    }
    \label{tab:scores}
  \end{table}

\begin{figure*}[!t]
    \centering
    \begin{subfigure}[t]{0.13\hsize}
      \centering
      \includegraphics[width=\columnwidth]{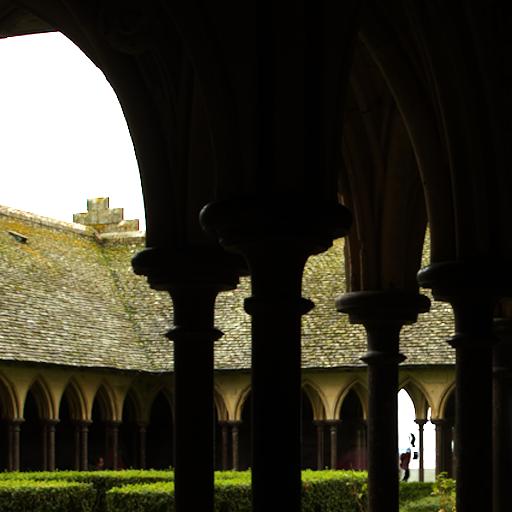}
      \caption{Input $x$\\
        TMQI: 0.7641\\
        Entropy: 3.1517\\
        NIQE: 5.5462\\
        BRISQUE: 35.702\\
        \label{fig:input5}}
    \end{subfigure}
    \begin{subfigure}[t]{0.13\hsize}
      \centering
      \includegraphics[width=\columnwidth]{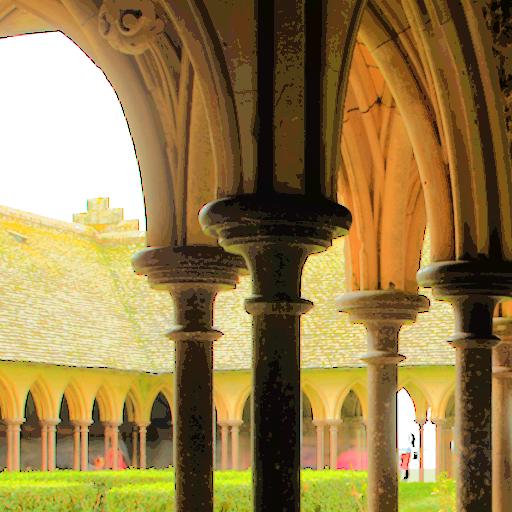}
      \caption{HE\\
        TMQI: 0.9320\\
        Entropy: 6.1185\\
        NIQE: 4.8412\\
        BRISQUE: 48.956\\
        \label{fig:he5}}
    \end{subfigure}
    \begin{subfigure}[t]{0.13\hsize}
      \centering
      \includegraphics[width=\columnwidth]{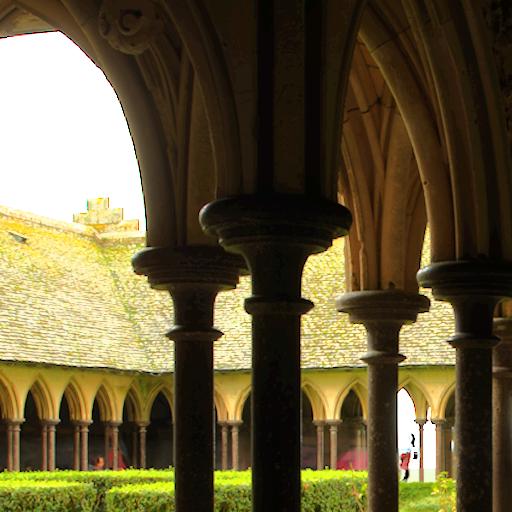}
      \caption{CACHE \cite{wu2017contrast}\\
        TMQI: 0.9157\\
        Entropy: 5.5691\\
        NIQE: 3.7613\\
        BRISQUE: 45.733\\
        \label{fig:cache5}}
    \end{subfigure}
    \begin{subfigure}[t]{0.13\hsize}
      \centering
      \includegraphics[width=\columnwidth]{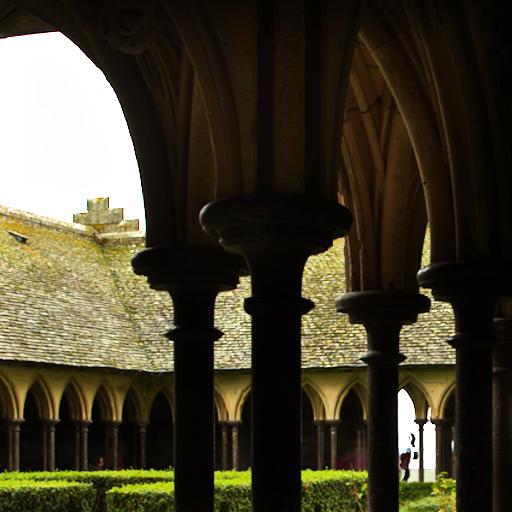}
      \caption{SRIE \cite{fu2016weighted}\\
        TMQI: 0.8235\\
        Entropy: 4.2474\\
        NIQE: 4.2855\\
        BRISQUE: 40.494\\
        \label{fig:srie5}}
    \end{subfigure}
    \begin{subfigure}[t]{0.13\hsize}
      \centering
      \includegraphics[width=\columnwidth]{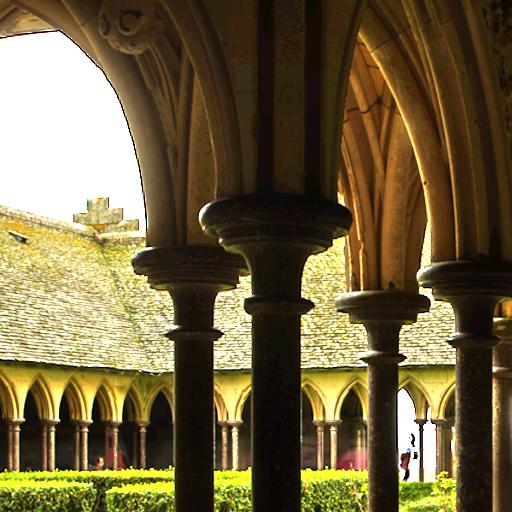}
      \caption{LIME \cite{guo2017lime}\\
        TMQI: 0.8940\\
        Entropy: 5.9848\\
        NIQE: 3.8960\\
        BRISQUE: 48.529\\
        \label{fig:lime5}}
    \end{subfigure}
    \begin{subfigure}[t]{0.13\hsize}
      \centering
      \includegraphics[width=\columnwidth]{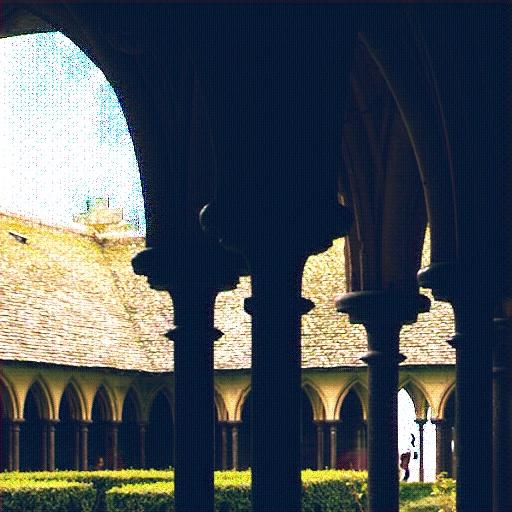}
      \caption{DRHT \cite{yang2018image}\\
        TMQI: 0.8170\\
        Entropy: 4.6590\\
        NIQE: 9.5757\\
        BRISQUE: 45.698\\
        \label{fig:drht5}}
    \end{subfigure}
    \begin{subfigure}[t]{0.13\hsize}
      \centering
      \includegraphics[width=\columnwidth]{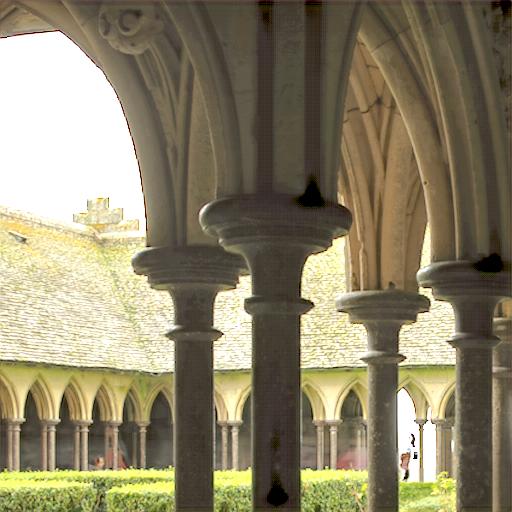}
      \caption{Proposed\\
        TMQI: 0.9351\\
        Entropy: 6.6459\\
        NIQE: 3.6894\\
        BRISQUE: 52.715\\
        \label{fig:proposed5}}
    \end{subfigure}
    \caption{Enhanced image results of six methods \label{fig:result5}}
  \end{figure*}

\begin{figure}[t]
    \centering
    \begin{subfigure}[t]{0.40\hsize}
      \centering
      \includegraphics[clip, width=\columnwidth]{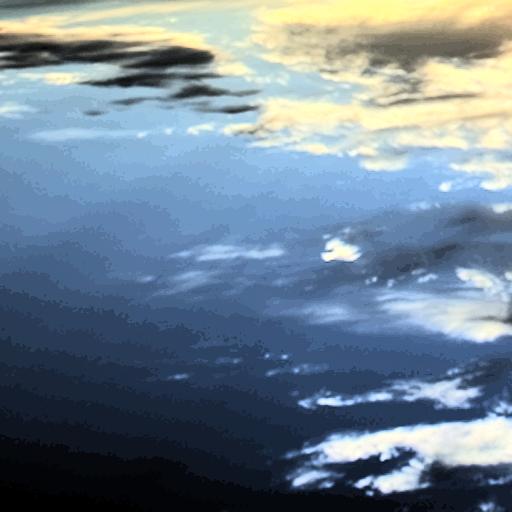}
      \caption{HE\\
        TMQI: 0.9238 \\
        Entropy: 5.9025\\
        NIQE: 5.4145\\
        BRISQUE: 44.950\\
        \label{fig:with_banding}}
    \end{subfigure}
    \begin{subfigure}[t]{0.40\hsize}
      \centering
      \includegraphics[clip, width=\columnwidth]{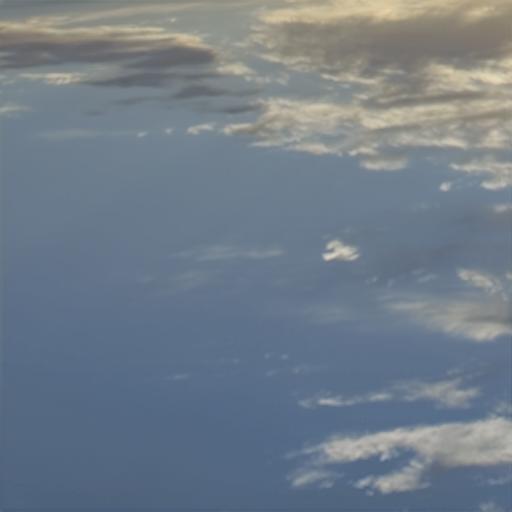}
      \caption{Proposed\\
        TMQI: 0.7688\\
        Entropy: 5.9650 \\
        NIQE: 4.4552\\
        BRISQUE: 45.805\\
      \label{fig:without_banding}}
    \end{subfigure}
    \caption{Banding artifacts in enhanced image \label{fig:banding}}
  \end{figure}

  Figure \ref{fig:result5} shows
  an example of the images enhanced by the six methods.
  From the figure,
  it is confirmed that the proposed method produced a higher-quality image
  that clearly represents dark areas in the image.
  The clarity of the image generated by DRHT is low
  due to the difficulty of generating an intermediate HDR image.
  In addition,
  the proposed method enhanced images without banding artifacts
  due to the quantized pixel values in the input images,
  but images enhanced by HE and CACHE include banding artifacts
  as shown in Fig. \ref{fig:banding}.
  This is because CNNs enhance images by taking into account
  local information around each pixel in images,
  whereas HE-based methods perform pixel-wise operations.
  The result denotes that the proposed method can not only enhance images,
  but also restore lost pixel values due to the quantization.

\begin{figure}[t]
    \centering
    \begin{subfigure}[t]{0.40\hsize}
      \centering
      \includegraphics[clip, width=\columnwidth]{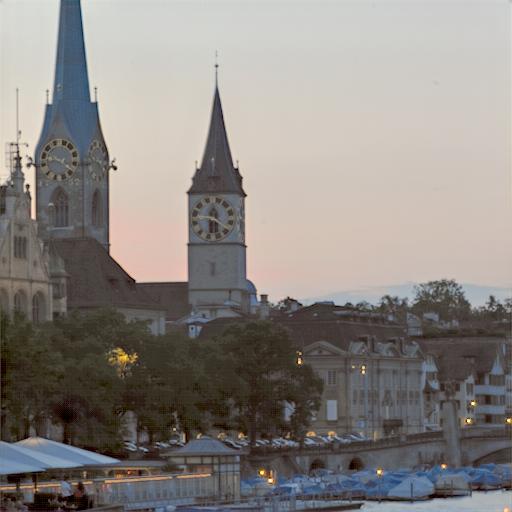}\\
      \caption{With global encoder.\\
        TMQI: 0.8488 \\
        Entropy: 6.6412\\
        NIQE: 2.6309\\
        BRISQUE: 50.297\\
        \label{fig:with_encoder}}
    \end{subfigure}
    \begin{subfigure}[t]{0.40\hsize}
      \centering
      \includegraphics[clip, width=\columnwidth]{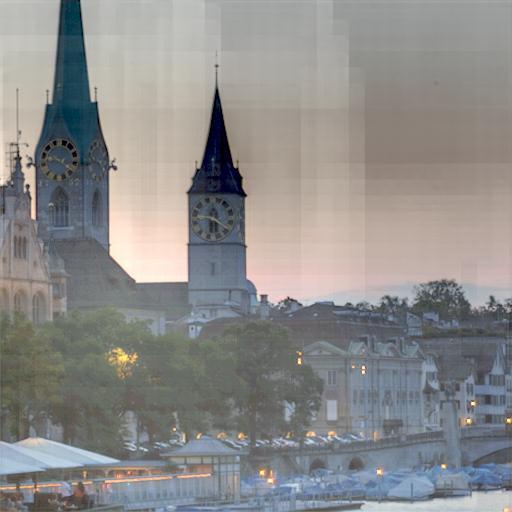}\\
      \caption{Without global encoder.\\
        TMQI: 0.7868\\
        Entropy: 7.2924\\
        NIQE: 2.5693\\
        BRISQUE: 57.115\\
      \label{fig:without_encoder}}
    \end{subfigure}\\
    \caption{Effectiveness of global encoder
      in proposed method.
      CNNs without global encoder create block-like artifacts in images.
      \label{fig:distortion}}
  \end{figure}

  Figure \ref{fig:distortion} illustrates that the effectiveness of the global encoder
  in the proposed network architecture.
  When the global encoder was not employed,
  CNNs caused images to be distorted, creating block-like artifacts.
  In contrast, the distortions were prevented
  when the global encoder was employed.
  Hence, the global encoder is effective to prevent distortions
  in enhanced images.

\begin{figure}[t]
    \centering
    \begin{subfigure}[t]{0.40\hsize}
      \centering
      \includegraphics[clip, width=\columnwidth]{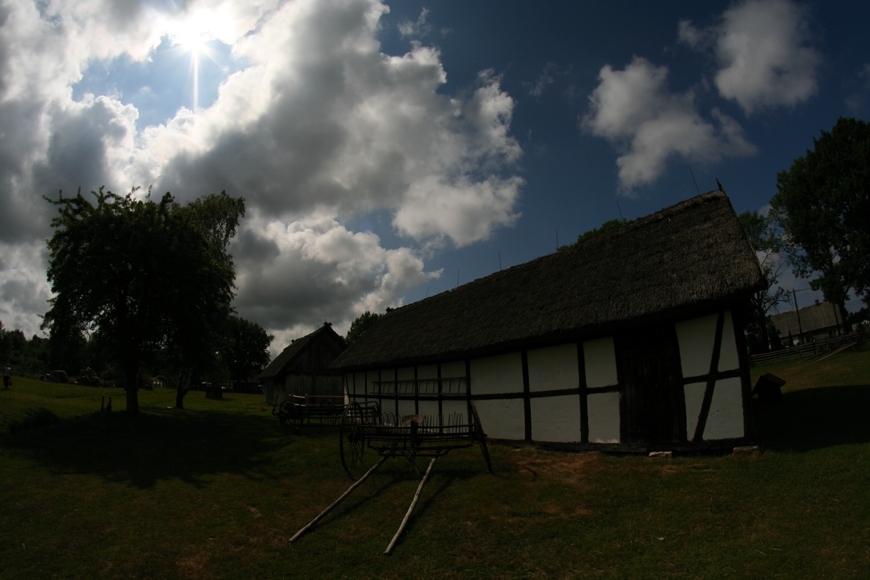}\\
      \caption{Input.\\
        Entropy: 5.0534\\
        NIQE: 1.6156\\
        BRISQUE: 46.1668\\
        \label{fig:real_input}}
    \end{subfigure}
    \begin{subfigure}[t]{0.40\hsize}
      \centering
      \includegraphics[clip, width=\columnwidth]{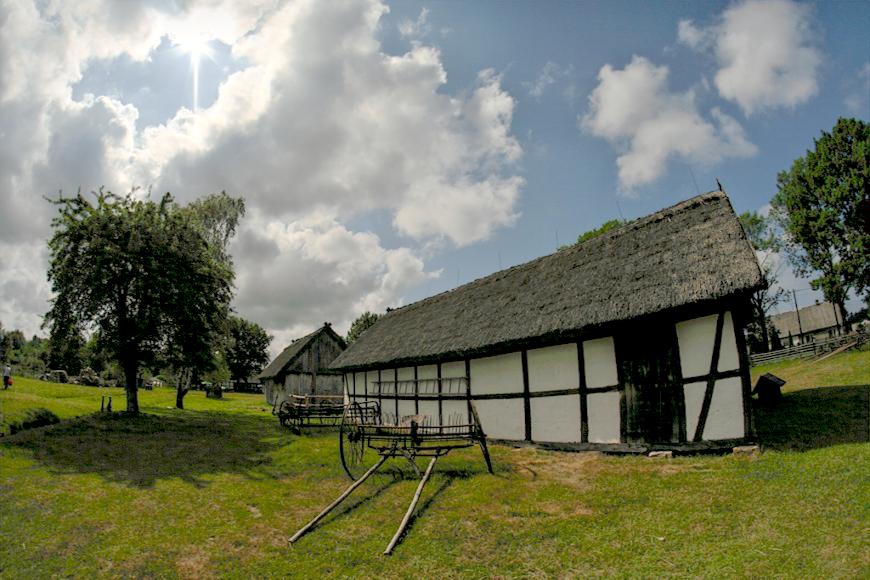}\\
      \caption{Proposed.\\
        Entropy: 7.2805\\
        NIQE: 1.7087\\
        BRISQUE: 50.9976\\
      \label{fig:real_proposed}}
    \end{subfigure}\\
    \caption{Enhanced image result from image that directly
      captured with digital camera.
      \label{fig:realdata}}
  \end{figure}

  An enhanced image result from an image that directly captured with a digital camera
  is shown in Fig. \ref{fig:realdata},
  where the input image was selected from \cite{easyhdr}.
  From the figure, it is confirmed that
  the proposed method is also effective for ordinary images,
  even when HDR images are only used to train CNNs.

  Those experimental results show that
  the proposed network architecture is effective for enhancing single images.

\section{Conclusion}
  In this paper, we proposed a novel CNN architecture considering
  both local and global features for image enhancement.
  The proposed architecture consists of a local encoder, a global encoder, and a decoder.
  The global encoder can effectively prevent distortions
  due to the lack of global image information required for image enhancement.
  Furthermore,
  the use of HDR images for training enables us to obtain
  higher-quality target images than those captured with cameras.
  Experimental results showed that the proposed method outperformed
  state-of-the-art conventional image enhancement methods including
  Retinex- and CNN-based methods, in terms of TMQI, entropy, NIQE, and BRISQUE.
  In addition, visual comparison results demonstrated that
  the proposed method can effectively enhance images without
  distortions such as banding and block artifacts.
  The performance of the proposed method depends on that of tone mapping.
  Hence, developing high-performance tone mapping operators
  will also be useful to image enhancement.

%

\end{document}